\documentclass[preprint,prd,nofootinbib,tightenlines,amsmath]{revtex4}
\usepackage{bbm}
\usepackage{amsfonts}
\usepackage{mathrsfs}
\usepackage{graphicx}
\usepackage{bm}

\begin{document}
\baselineskip=15pt \parskip=5pt

\vspace*{3em}

\title{Study of a WIMP dark matter model with the updated
results of CDMS II}

\author{Lian-Bao Jia}

\author{Xue-Qian Li}
\email{lixq@nankai.edu.cn}

\affiliation{Department of Physics, Nankai University, Tianjin
300071, China
\\}

\begin{abstract}

The new observation of CDMS II favors low mass WIMPs. Taking the
CDMS II new results as inputs, we consider a SM singlet: the darkon
as the dark matter candidate, which can be either scalar, fermion or
vector. It is found that the simplest scenario of DM+SM conflicts
with the stringent constraint set by the LHC data. New physics
beyond the SM is needed, and in this work, we discuss an extended
standard model $SU_L(2)\otimes U_Y(1)\otimes U(1)'$ where $U(1)'$
only couples to the darkon. The new gauge symmetry is broken into
$U_{em}(1)$ and two neutral bosons $Z^0$ and $Z'$ which are mixtures
of $W^3_{\mu},\; B_{\mu}\; X_{\mu}$ are resulted in. Following the
literature and based on the CDMS data, we make a complete analysis
to testify the validity of the model. The cross section of the
elastic scattering between darkon and nucleon is calculated, and the
DM relic density is evaluated in the extended scenario as well. It
is found that considering the constraints from both cosmology and
collider experiments, only if $Z'$ is lighter than $Z^0$, one can
reconcile all the presently available data.

\end{abstract}

\maketitle

\section{Introduction}

Recently, the CDMS Collaboration reports that three WIMP-candidate
events were observed~\cite{Agnese:2013rvf} by using the silicon
detectors. With a final surface-event background estimate of
$0.41^{+0.20}_{-0.08}(stat.)^{+0.28}_{-0.24}(syst.)$, they indicate
that the highest likelihood occurs for a WIMP mass of 8.6 GeV/$c^2$
and spin-independent WIMP-nucleon cross section of
$1.9\times10^{-41}$ cm$^2$. This observation seems to contradict
with the results of XENON100 \cite{Aprile:2012nq}. Hooper
\cite{Hooper:2013cwa} re-analyzed the data of XENON100 and reached a
different conclusion, namely the two experimental results can be
reconciled. Therefore in this work, we take the CDMS results as
inputs to study the dark matter. We will test a viable model
proposed in literature, namely check whether both the astronomical
observation and constraints from the collider experiments can be
simultaneously satisfied in this scenario.

As is well known, none of the standard model (SM) particles can meet
the criterion to stand as dark matter (DM) candidates. Many
particles beyond the SM are proposed, for example the primordial
black holes, axions, heavy neutrinos, the lightest supersymmetric
neutralino, etc. Among them, the darkon model namely a SM singlet
scalar
\cite{Silveira:1985rk,McDonald:1993e,Burgess:2000yq,He:2007tt,He:2008qm,Mambrini:2011ik}
which interacts with the SM particles by exchanging Higgs boson
only, probably is the simplest version for the dark matter
candidates. The spin-independent cross section for the
darkon-nucleon elastic scattering might be measured by the earth
detectors. The typical recoil energy is $\Delta E_R\sim (\mu
v)^2/m_A$, where $\mu$ is dark matter-nucleus reduced mass, $v$ is
DM velocity, and $m_A$ is target nucleus mass. The WIMPs with not
very heavy masses will weaken the bounds in detector search, and the
low mass WIMPs (mass around 10 GeV) are more of our concern in this
work.

Thanks to the successful operation of LHC where the Higgs boson
signals have been observed \cite{Aad:2012tfa,Chatrchyan:2012ufa}, it
provides a possible means to directly detect the dark matter
particles on the earth if they indeed exist. It means that all the
proposed dark matter candidates and possible new interactions by
which the DM particles interfere with our detector would withstand
the stringent test on the earth colliders. Namely, if the proposed
DM particles, especially the lighter ones, are not observed at LHC
as expected, the concerned model fails or needs to be modified. As
indicates in~\cite{He:2007tt}, if the mass of the darkon is lighter
than half of Higgs mass, Higgs would decay into a darkon pair which
is a channel with invisible final products, and the simplest version
of scalar darkon+SM may fail. That is to say, if darkon's coupling
to Higgs is not much smaller than 1, a large partial width is
expected and it obviously contradicts to the measured value of the
invisible width of the SM Higgs. As a possible extension of the scalar
darkon+SM version, the two-Higgs-doublet model was discussed
in~\cite{He:2007tt,He:2008qm,He:2013suk} and there seems to be a large
parameter space to accommodate both the LHC data on Higgs and the
CDMS observations.

We also find that the scenario of darkon+SM, no matter the darkon is
scalar, fermionic or vector, definitely fails, thus a new
interaction beyond the SM is needed. Alternatively, we propose an
interaction beyond the SM as the darkon+SM+an extra $U(1)'$. The
extended gauge group $SU_L(2)\otimes U_Y(1)\otimes U(1)'$ breaks
into $U_{em}(1)$ and two neutral bosons $Z^0$ and $Z'$ are resulted
in. $Z^0$ and $Z'$ are mixtures of $W^3_{\mu},\; B_{\mu}$ and
$X_{\mu}$ which is the gauge boson of the newly introduced $U(1)'$,
while the photon remains massless. In this scenario, to be
consistent with the CDMS and LHC data simultaneously, we should
assume that the coupling between the darkon and Higgs boson to be
very small and the interaction by exchanging Higgs between the
detector material and darkon can be safely ignored. Therefore, the
possible decays of Higgs boson into darkons are almost forbidden and
one cannot expect to measure the mode at LHC at all. The scattering
between the darkon and nucleons is due to exchanging the gauge boson
$Z^0$ and $Z'$. Definitely, such interaction may also exist in the
decays of quarkonia, i.e. if the measurement of heavy quarkonia,
such as bottomonia, are very precise, one may observe their decays
into invisible final products besides the SM neutrino-anti-neutrino
pairs. But it is estimated that the branching ratios for such decays
of heavy quarkonia are too small to be reliably measured in any of
our present facilities. Besides, when the bottomonia are lighter
than the new invisible final products, these decays are also
forbidden. Therefore this proposed darkon+SM+$U(1)'$ is safe with
respect to the present experimental constraints. Moreover, the
observed relic density of dark matter in our universe sets one more
constraint on our model parameter space.

This work is organized as follows. After this introduction, we first
consider the simple version of scalar, fermionic, and vector darkon
within the framework of standard model plus darkon, then we derive
the formulas of the cross section between nucleon and darkon, as
well as the decay width of Higgs into invisible darkons. We further
derive the corresponding formulas for the aforementioned extended
version darkon+SM+$U(1)'$. Then in the following section, we
numerically evaluate the cross sections of darkon-nucleon elastic
scattering with the two scenarios. We indicate that the simple
version does not satisfy the constraint set by the LHC data as long
as we take the CDMS data as inputs, but in the extended version
there is a parameter space to accommodate both the experimental
measurements. The last section is devoted to our brief summary and
discussions.

\section{Darkon+SM}

\begin{figure}[t]
\includegraphics[bb=70 545 525 770,width=3in]{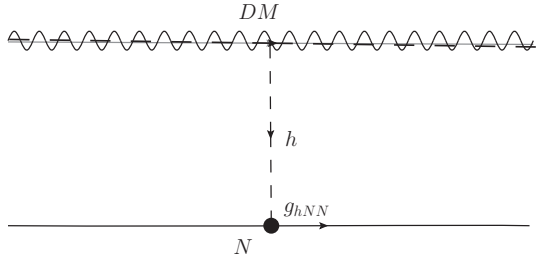}
\caption{The elastic scattering between dark matter and nucleon with
Higgs boson exchanged.\label{Fig-scatt}}
\end{figure}

In this work, as CDMS data suggested, we focus on low mass WIMPs.
The WIMP particle could be an $SU_c(3)\times SU_L(2)\times U_Y(1)$
singlet, i.e. either a scalar or fermion or vector darkon
\cite{He:2007tt,Lebedev:2011iq,Djouadi:2011aa}. In the scenario of
darkon+SM, the elastic scatting between darkon and the detector
material is realized via t-channel Higgs exchange, as described in
Fig.~\ref{Fig-scatt}.

\subsection{Scalar darkon}

Let us consider a scalar type WIMP DM, namely a scalar darkon first.
This type DM has been discussed in~\cite{He:2007tt}, and here for
completeness we first repeat some relevant procedures. The
Lagrangian is written
as~\cite{Silveira:1985rk,McDonald:1993e,Burgess:2000yq,He:2007tt}
\begin{eqnarray}  \label{eqs1}
{\cal L}&=&{\cal L_{SM}}
-\frac{\lambda_S}{4}\,S^4+\frac{1}{2}\partial^\mu
S\,\partial_\mu^{}S\ - \frac{m_0^2}{2}\,S^2 - \lambda\,
S^2\,H^\dagger H~.
\end{eqnarray}
Here $\lambda_S^{}$, $m_0^{}$, and $\lambda$ are free parameters to
be determined by fitting data. It has been indicated in earlier
works that the scalar darkon field has no mixing with the Higgs
field, and this can avoid fast decaying into SM particles because
dark matter particles must be sufficiently stable and survive from
the Big Bang to today. From Eq. (\ref{eqs1}), the SM singlet scalar
darkon can be further written as
\begin{eqnarray}  \label{eqs2}
{\cal L_{S}}&=&-\frac{\lambda_S}{4}\,S^4+\frac{1}{2}\partial^\mu
S\,\partial_\mu^{}S\ - \frac{m_0^2+\lambda\,v^2}{2}\,S^2 -
\frac{1}{2}\lambda\, S^2\,h^2-\lambda\,vS^2\,h~.
\end{eqnarray}
The Higgs-nucleon coupling $g_{hNN}$ is needed in calculating the
scatting process, ${\cal L}_{hNN}^{}=$ $-g_{hNN}\,\bar NN\,h$. Here
we adopt the value of $g_{hNN}$ given by He et al.~\cite{He:2008qm},
\begin{eqnarray} \label{eqs3}
g_{hNN}\,\bar NN=<N|\frac{k_{u}}{v}(m_{u}\bar uu+m_{c}\bar
cc+m_{t}\bar tt)+\frac{k_{d}}{v}(m_{d}\bar dd+m_{s}\bar ss+m_{b}\bar
bb)|N>~,
\end{eqnarray}
and $g_{hNN}\simeq1.71\times10^{-3}$.  The cross-section of scalar
DM-nucleon elastic scatting is
\cite{Silveira:1985rk,McDonald:1993e,Burgess:2000yq}
\begin{eqnarray} \label{s-cs}
\sigma_{\rm el}^{} \,\,\simeq\,\,
\frac{\lambda^2\,v^2\,g_{hNN}^2\,m_N^2}{\pi\,(p_{D}+p_{N})^2\,
m_h^4} \,\,.
\end{eqnarray}
Here, $p_{D}$, $p_{N}$ are the momenta of the initial DM and
nucleon. For low energy elastic scatting,
$(p_{D}+p_{N})^2\simeq(m_{D}+m_N)^2$, and $m_{D},~m_N$ are masses of
DM, nucleon respectively. Substituting the darkon mass 8.6 GeV and
the cross section $1.9\times10^{-41}$ cm$^2$ as given by CDMS II
into the above formula (\ref{s-cs}), we can fix the effective
coupling of Higgs-darkon.

The Higgs signals have been observed at
LHC~\cite{Aad:2012tfa,Chatrchyan:2012ufa} and $m_h=$ 125 GeV, so by
the data of CDMS, $\lambda\approx0.148$ is determined. The partial
width of Higgs decaying into two scalar darkons is
\begin{eqnarray} \label{hss}
\Gamma_{h\rightarrow{SS}}=\frac{\lambda^2v^2}{8\pi{m_h}}\sqrt{1-\frac{4m_{D}^2}{m_h^2}}~.
\end{eqnarray}
Substituting Higgs mass into the equation,
$\Gamma_{h\rightarrow{SS}}\approx0.418$ GeV is obtained. The main
decay channel in SM is $h\rightarrow{b\bar b}$. In the Born
approximation, the width of this channel is
\cite{Resnick:1973vg,Ellis:1975ap}
\begin{eqnarray} \label{hbb}
\Gamma_{Born}(h\rightarrow{b\bar b} )=\frac{3 G_{F}}{4\sqrt{2}\pi}
M_h m_b^2 \beta_b^3~.
\end{eqnarray}
Here $\beta=\sqrt{1-4m_b^2/M_h^2}$, and $G_{F}$ is the Fermi
coupling constant. With $G_{F}$=1.166$\times10^{-5}~GeV^{-2}$ and
$m_b(\overline{MS})=4.18~GeV$, we can obtain
$\Gamma_{Born}(h\rightarrow{b\bar b} )\approx0.00427$ GeV. Thus the
branching ratio ${B}_{h}\to invisible$  would be too large.

\subsection{Fermionic and vectorial darkons}

In the spin-$\frac{1}{2}$ darkon+SM scenario, the effective
interaction can be written as
\begin{eqnarray} \label{hs1/2}
{\cal L}_ {int}=-\lambda\bar \psi_{D}\psi_{D} h~.
\end{eqnarray}
The  cross-section of the low energy fermion-darkon-nucleon elastic
scatting is
\begin{eqnarray} \label{css1/2}
\sigma_{\rm el}^{} \,\,\simeq\,\, \frac{\lambda^2\,m_D^2
 g_{hNN}^2\,m_N^2}{\pi\,(p_{D}+p_{N})^2\, m_h^4} \,\,.
\end{eqnarray}
The partial width of Higgs decaying into two darkon spinors is
\begin{eqnarray} \label{hss}
\Gamma_{h\rightarrow{DM}}=\frac{\lambda^2
{m_h}}{8\pi}\sqrt{1-\frac{4m_{D}^2}{m_h^2}}~.
\end{eqnarray}
In this case, the invisible decay width is unbearably large when
$\lambda$ is at order of unity. It means that such spin-${1\over 2}$
darkon+SM scenario must also be abandoned.

For a vector darkon, the effective lagrangian can be written as
\begin{eqnarray} \label{hs1}
{\cal L}_ {VH}={\lambda} V^{\mu}V_{\mu} H^{\dag}H~.
\end{eqnarray}
The cross section of the elastic scattering between a vector darkon
and nucleon via a Higgs boson exchange is
\begin{eqnarray} \label{cses}
\sigma_{\rm el}^{} \,\,\simeq\,\,
\frac{\lambda^2\,v^2\,g_{hNN}^2\,m_N^2}{\pi\,(p_{D}+p_{N})^2\,
m_h^4}~.
\end{eqnarray}

The numerical results for the vector darkon are similar to the two
above cases for scalar and fermion darkons, namely with the darkons
possessing a low mass of order of 10 GeV and the spin-independent
cross section as determined by the CDMS data, the partial width of
Higgs decaying into invisible final products would be too large to
be tolerated.

The above results indicate that  the simplest scenario of darkon+SM,
no matter the SM singlet darkon is a scalar, fermion or vector,
cannot reconcile the cosmological observation of CDMS and the LHC
data. Then one should invoke an extended version of SM i.e. a
darkon+SM+BSM scenario. But what model beyond standard model (BSM)
which can be applied to explain the CDMS observation and the LHC
data simultaneously, is a problem. There are many different
proposals, and below we will investigate a naturally extended
version of the SM, i.e. introducing an extra $U(1)'$ gauge field
which would be broken and a new vector boson $Z'$ is induced.

\section{Darkon+SM+$U(1)'$}

For low mass darkon model, the simple version darkon+SM where
darkons interact with the SM particles in detector by exchanging
Higgs boson at t-channel, definitely fails to reconcile the
observation of CDMS and LHC data, and therefore needs to be
modified. To tolerate the CDMS and LHC data, besides the
two-Higgs-doublet model mentioned above, alternatively, for example, the
sneutrino dark matter which interacts with the detector material
dominantly via exchanging SM Z-boson, was discussed
in~\cite{Choi:2013fva}.

In this work, we would study the effects of
an extended SM by adding an extra
$U(1)'$~\cite{Brahm:1989jh,deCarlos:1997yv,Erler:2002pr,Kors:2004dx,Barger:2004bz,Kors:2005uz,Langacker:2008yv}
which only interacts with the darkons (no matter scalar, fermion or vector darkons)
into the gauge group, as $SU_L(2)\otimes U_Y(1)\otimes U(1)'$ whose
gauge bosons are respectively $W^{\pm}_{\mu},\; W^3_{\mu},\;
B_{\mu}$ and $X_{\mu}$ (for more discussions about this model, see
e.g.,~\cite{Kumar:2006gm,Chang:2006fp,Feldman:2007wj,Mambrini:2010dq,Frandsen:2011cg}).
The extended symmetry later breaks into $U_{em}(1)$. As a
consequence, besides the regular charged $W^{\pm}$, two neutral
gauge bosons $Z^0$ and $Z'$ gain masses after the symmetry breaking
while the photon remains massless.

It is noted that a small mixing between the SM $Z$ and $X$ results
in the physical mass eigenstates $Z^0$ and $Z'$. Since the mixing is
required to be very small the resultant $Z^0=\cos\varphi Z +
\sin\varphi X$ is almost the SM Z boson whereas $Z'$ is
overwhelmingly dominated by $X$. Concretely, after $SU(2)_L\times
U(1)_Y\times U(1)'$ breaking, one has
\begin{eqnarray} \label{gauge1}
\left (
\begin{array}{c}
  A_\mu \\
  Z_\mu^0 \\
  Z_{\mu}'
\end{array}
\right )=\left [
\begin{array}{ccc}
           \cos\theta_w & \sin\theta_w & 0 \\
           -\sin\theta_w\cos\varphi & \cos\varphi\cos\theta_w & \sin\varphi \\
           \sin\theta_w\sin\varphi & -\sin\varphi\cos\theta_w & \cos\varphi
         \end{array} \right ]
         \left (
\begin{array}{c}
  B_\mu \\
  W_\mu^3 \\
  X_{\mu}
\end{array}
\right ).
\end{eqnarray}

Assuming $X_{\mu}$ of $U(1)'$ only couples to the darkon but not the
SM particle whereas $Z_{\mu}$ only couples to the SM particles, thus
the interaction between the darkon and SM particles must be realized
via the small mixing. Namely the effective interaction amplitude
between the darkon and protons or neutrons in the earth detector
must be proportional to $\sin\varphi\cdot\cos\varphi$. To be
consistent with experiments, $\varphi$ should be very small, i.e.
$\sin\varphi\ll1,\,\cos\varphi\sim1$.

Since the new effective vertex $V_{DV(A)D}$ is a coupling between
scalar, fermionic or vector darkon with the gauge boson, the Lorentz
structure are well determined even though the coupling constants
might be model dependent. Feytsis and Ligeti \cite{Freytsis:2010ne}
listed all possible operators and indicated which one(s) is
suppressed by $q^2$ or $v^2$ where $q$ is the exchanged momentum and
$v$ is the speed of the dark matter relative to the earth detector.
Thus, in this work, we only concern the unsuppressed
spin-independent scattering processes which may correspond to the
recently observed events. Below, we will be focusing on the
fermionic darkon and give all the details, but for completeness we
also briefly discuss the cases for the scalar and vector darkons.

\subsection{Fermionic darkon}

Let us consider the fermionic darkon first. The axial-vector
component of the gauge boson may induce a fermionic darkon-nucleon
interaction which is not suppressed by $q^2$ or $v^2$, even though
this coupling would result in a spin-dependent cross section
\cite{Freytsis:2010ne}. For easily handling, here we consider a
right-handed darkon with the vertex
$i\lambda\gamma^{\mu}\frac{1+\gamma^5}{2}$ to interact with the SM
particles via exchanging Z-boson. The darkon-nucleon elastic
scattering cross section is calculated for two cases: $m_{Z'} \gg
m_{Z^0}$ and $m_{Z'} \ll m_{Z^0}$ respectively, and corresponding DM
relic density is also computed.

\subsubsection{The case $m_{Z'}\gg m_{Z^0}$}

In this case, the darkon-nucleon elastic scattering occurs mainly
via exchanging $Z^0$. The fugacity speed of WIMP is about
$220\sim544$ km/s \cite{Smith:2006ym}. For the low energy
$Z^0$-nucleon interaction, the hadronic matrix element can be
expressed as ~\cite{Garvey:1993sg,Garvey:1995im,Alberico:2001sd}
\begin{eqnarray}
\langle p',s'\mid J_\mu^{Z^0}\mid
p,s\rangle=\sqrt{\frac{G_F}{\sqrt{2}}}\overline{\mathcal
{U}}_N(p',s')\left [ G_A^z \gamma_\mu\gamma^5+F_1^z \gamma_\mu+F_2^z
\frac{i\sigma_{\mu\nu }q^{\nu}}{2M_N}\right ] \mathcal {U}_N(p,s)~.
\end{eqnarray}
Here $\mathcal {U}_N$, $M_N$, $q$ are the nucleon's wave function,
mass, and momentum transfer respectively. $G_A^z,~F_1^z,~$ and
$F_2^z$ are the relevant form factors. Those form factors can be
determined by the data of elastic scattering between neutrino and
nucleon, since for this neutral current scattering process only
$Z^0$ exchange is dominant (the new boson $Z'$ is suppressed by a
factor $\sin^4\varphi$ in this process).

Here we adopt the way given in Ref. \cite{Garvey:1993sg,Garvey:1995im} to
define the form factors. In the form of quark currents, the
hadronic matrix element is written as
\begin{eqnarray}
\langle p',s'\mid J_\mu^{Z^0}\mid
p,s\rangle=\sqrt{\frac{G_F}{\sqrt{2}}}\overline{\mathcal
{U}}_N(p',s')\sum_i\left [ \overline{q}_i
\gamma_\mu(1-\gamma^5)t_z{q}_i-2Q_i \sin^2\theta_w \overline{q}_i
\gamma_\mu{q}_i \right ] \mathcal {U}_N(p,s).
\end{eqnarray}
The form factors are written as
\begin{eqnarray}
&G_A^z=-\frac{G_A^3\tau_3}{2}+\frac{G_A^s}{2},\\
&F_1^z=(1-2\sin^2\theta_w)F_1^3\tau_3-2\sin^2\theta_wF_1^1-\frac{F_1^s}{2},\\
&F_2^z=(1-2\sin^2\theta_w)F_2^3\tau_3-2\sin^2\theta_wF_2^1-\frac{F_2^s}{2},
\end{eqnarray}
where the isospin factor $\tau_3=$ +($-$) for proton (neutron), and
\begin{eqnarray}
F_j^1=\frac{F_j^p+F_j^n}{2},\\
 F_j^3=\frac{F_j^p-F_j^n}{2},
\end{eqnarray}
with j=1,2.

Defining $Q^2=-q^2$, since $Q^2/m_N^2\ll1$, for darkon-nucleon
scattering via exchanging $Z^0$ boson, we can set the values of the
form factors at $Q^2=0$. At $Q^2=0$,
$F_1^p=1,~F_1^n=0,F_2^p=1.7928,~F_2^n=-1.9130$ \cite{Garvey:1995im}.
In the limit of $Q^2=0$, the parameters corresponding to the strange
part are $G_1^s(0)=\Delta s$, $F_1^s(0)=0$,
$F_2^s(0)=\mu_s$~\cite{Garvey:1995im,Alberico:2001sd,Forkel:1995ff},
and here we take the fitted results
$G_1^s(0)=-0.15\pm0.07,~F_1^s(0)=0,~F_2^s(0)=0,~M_A=1.049\pm0.019$,
($\chi^2=9.73$ at 13 DOF)~\cite{Garvey:1995im,Alberico:2001sd}. The
PDG average value of $G_A^3$ is $G_A^3=1.2701\pm0.0025$
~\cite{Beringer:1900zz}. So at $Q^2=0$, the form factors are
$G_A^z\approx-0.710~(0.560)$ for proton (neutron), and $F_1^z=$
$0.5-2\sin^2\theta_w$ ($-0.5$) for proton (neutron). The
contribution from $F_2^z$ term is suppressed at $Q^2=0$. If
considering the conservation of the vector currents and just using
the valence quarks in the nucleon, the same result can be obtained
for the vector form factor $F_1^z$.

As the darkon is non-relativistic, in the limit
$\frac{P^\mu}{m}\rightarrow(1,\epsilon)$, the darkon-nucleon elastic
scattering cross section with $Z^0$ exchanged at t-channel can be
written as
\begin{eqnarray}
\sigma_{\rm el} \simeq \frac{\sqrt{2}
G_F\lambda^2\,\sin^2\varphi\,m_{D}^2\,m_N^2\,(3\,{G_A^z}^2+{F_1^z}^2)}{4\pi\,(p_{D}+p_{N})^2\,
m_{Z^0}^2} \,\,.
\end{eqnarray}
It is noted that $G_A^z$ is spin-dependent (SD), and $F_1^z$ is
spin-independent (SI). For large mass target nuclei, such as the
silicon, germanium and xenon targets, the spin-independent
interaction is enhanced by the atomic number $A^2$ (but not exactly,
see below for details) in the target nucleus, so the
spin-independent interaction is more sensitive than the
spin-dependent case, as discussed in Ref.~\cite{Freytsis:2010ne}.
Thus we can drop the spin-dependent term $G_A^z$ but just keep the
spin-independent term $F_1^z$ for large mass target nuclei
scattering. For proton, $F_1^z(p)=$
$0.5-2\sin^2\theta_w\approx0.038$, while for neutron
$F_1^z(n)=-0.5$. Thus, the darkon-neutron scattering is dominant and
the scattering cross section of darkon-nucleus via exchanging a
neutral gauge boson $Z$ should be proportional to $(A-Z)^2$ instead
of $A^2$. Thus a factor of about 0.25 might exist and when analyzing
the data to extract the information about the dark matter-nucleon
interaction, this factor should be considered.

Substituting the CDMS
II results for darkon-neutron elastic scattering: $m_D \sim 8.6$
GeV/$c^2$ and the elastic cross section $\sigma_{\rm el} \sim
4\times1.9\times10^{-41}$ cm$^2$ into the relevant formulas, we obtain
$\lambda^2\,\sin^2\varphi\approx6.88\times10^{-3}$. To require the
coupling constant $\alpha_D=\frac{\lambda^2}{4\pi}<1$, the upper
limit of $\lambda$ is $\sqrt{4\pi}$.

In fact, the LEP data set a stringent constraint on the coupling and
mixing.  The width of $Z^0$ decaying into invisible products is
$\Gamma(invisible)=499.0\pm1.5$ MeV~\cite{Beringer:1900zz}. It is
assumed in our scenario, that subtracting the main contribution of
neutrinos from the measured width, the rest can be attributed to the
darkon products. Thus  we can use the data to estimate the range of
$\varphi$ with some unavoidable uncertainties. The width of $Z^0$
decaying into a darkon pair is formulated as
\begin{eqnarray}
\sin^2\varphi \Gamma_{D}=\frac{\lambda^2 \sin^2\varphi
(m_{Z^0}^2-m_D^2)}{24 \pi m_{Z^0}}\sqrt{1-\frac{4
m_D^2}{m_{Z^0}^2}}~.\label{widthZ}
\end{eqnarray}
Then the total width of $Z^0$ decaying into invisible products is
\begin{eqnarray}
\cos^2\varphi \Gamma_{\nu\bar{\nu}}+\sin^2\varphi
\Gamma_{D}\leq\Gamma_{\nu\bar{\nu}}+\sin^2\varphi
\Gamma_{D}\approx505.7~MeV,\label{TotalZ}
\end{eqnarray}
and this value is larger than the experimentally measured value (the
central value) for invisible products. If the mixing angle
$\sin^2\varphi$ is reduced to an order of 0.01, this could satisfy
LEP data. However, this mixing angle is too large to be accepted
because the SM electroweak sector would be seriously affected to
conflict with all the previous well-done measurements.

Another constraint  comes from the observed density of dark
matter in our space.

The motion of the darkon is non-relativistic, the invariant mass of
a darkon pair can be approximated as $\sqrt s\simeq2m_D$ where $m_D$
is the darkon mass. In order to get the DM relic density, we need to
calculate DM annihilation cross section. In
Ref~\cite{Burgess:2000yq}, the scalar-mediated (Higgs) 2$\to$2
annihilation cross-section of DM pair into SM particles is given.
But as discussed in the introduction, we choose an alternative
scenario where the coupling of Higgs boson with darkon is too small
to make any substantial contributions to the darkon-nucleon
scattering and as well as the dark matter annihilation.

Here, the annihilation cross section of darkons is dominated by the
process that a darkon pair annihilates into a virtual gauge boson
($Z^0$ or $Z'$) which later transits into SM final states. Considering
the case that the intermediate boson has a narrow width compared
with its mass at the pole, the cross section is written as
\begin{eqnarray}\label{cs-general}
\sigma_{\rm ann}\,&=& \frac{1}{2}\sigma_{\rm
ann}^{Dirac}=\frac{1}{2}\frac{1}{\beta_i (2s_1+1)(2s_2+1)}\frac{
\lambda^2 \sin^2\varphi \cos^2\varphi}{\bigl(s-M^2)^2+M^2 \Gamma^2} \nonumber \\
&&\times[2(s-m_D^2)\frac{\tilde{\Gamma}_f}{\sqrt{s}}+
(\frac{s}{M^2}-1)^2 \frac{m_{Z^0}^2 G_F}{2\sqrt{2}} \frac{N_c
\beta_f c_a^2 m_f^2 m_D^2}{\pi s }].
\end{eqnarray}
A factor $\frac{1}{2}$ appears for fermion dark matter which is composed of
particle and anti-particle simultaneously, and the annihilation only occurs between
particle and its anti-particle (similarly, the factor
$\frac{1}{2}$ exists for complex scalar DM, while this factor is equal
to 1 for real scalar, Majorana fermion DM). $s=(p_1+p_2)^2$, $M$ is
the mass of the intermediate boson, and $\Gamma$ is the total width
of the intermediate boson. $s_1,s_2$ are the darkon spin
projections. $\tilde{\Gamma}_f$ is the rate of the virtual boson
transiting into SM fermions (quarks or leptons), to obtain it, one
only needs to replace the intermediate boson mass by $\sqrt s$ in
the calculations. $N_c$ is the color factor. $c_a$ is the axial
vector current parameter, here $c_a^2=1$.
$\beta_i=\sqrt{1-4m_D^2/s}$, ${\beta}_f=\sqrt{1-4m_f^2/s}$ are the
kinematic factors.

In the case of $m_{Z'}\gg m_{Z^0}$, the annihilation of a darkon
pair into SM particles is dominated by $Z^0$ with the mixing
component, namely via $darkon+darkon \to {Z^0} \to SM $. Using
formula (\ref {cs-general}), we can get the annihilation cross
section. The DM relic density $\Omega_D$ is determined by the
thermal dynamics of the big-bang cosmology. The approximate values
of the relic density and freeze-out temperature are
~\cite{Kolb:1990vq,Griest:1990kh}
\begin{eqnarray}\label{relicD}
\Omega_D h^2 \,\simeq\, \frac{1.07\times 10^9\,GeV^{-1}\, x_f}{
\sqrt{g_*}\, m_{\rm Pl}\,\langle\sigma_{\rm ann} v_{\rm rel}
\rangle\rm},
\end{eqnarray}
\begin{eqnarray}\label{xf}
x_f^{} \,\simeq\, \ln\frac{0.038\,g\,m_{\rm
Pl}\,m_D\,\langle\sigma_{\rm ann}v_{\rm rel}\rangle}{ \sqrt{g_*\,
x_f^{}}}.
\end{eqnarray}
Here $h$ is the Hubble constant in units of 100 km/(s $\cdot$ Mpc),
and $m_{\rm Pl}^{}=1.22\times10^{19}$ GeV is the Planck mass.
$x_f^{}=m_D^{}/T_f^{}$\, with $T_f^{}$ being the freezing
temperature, $g_*^{}$ is the number of relativistic degrees of
freedom with masses less than $T_f^{}$. $\langle\sigma_{\rm
ann}^{}v_{\rm rel}^{}\rangle$ is the thermal average of the
annihilation cross section of DM pair transiting into SM particles
and $v$ is the relative speed of the DM pair in their center of mass
frame, and $g$ is the number of degrees of freedom of DM. In this
work, the DM particle is assumed to be the darkon, which we describe
above. The thermal average of the effective cross section
is~\cite{Gondolo:1990dk}
\begin{eqnarray}
\langle \sigma_{ann}^{} v_{rel} \rangle=\frac{1}{8 m_D^4\, T
K_2^2(\frac{m_D}{T})} \int_{4m_D^2}^{\infty}
ds\,\sigma_{ann}^{}\,\sqrt{s}(s-4m_D^2) K_1(\frac{\sqrt{s}}{T})~,
\end{eqnarray}
where $K_i(x)$ is the modified Bessel functions of order $i$.

We calculate the cross section of low mass darkon pairs (the mass of
darkon is supposed to be of order 10 GeV) annihilating into SM
leptons and quarks (except top quark) via  $Z^0$ exchanged. $x_f^{}$
is obtained by solving the Eq. (\ref{xf}) iteratively. The effective
degrees of freedom $g_\ast$ is varying with the freeze-out
temperature $T_f^{}$, and we take the data of Gondolo-Gelmini
effective degrees of freedom in MicrOMEGAs 3.1 at $T_{QCD}=$ 150 MeV
\cite{Belanger:2013oya}. For $m_D \sim 8.6$ GeV, the DM density is
$\Omega_D^{} h^2\approx0.593$. The current PDG value for cold DM
density is \,$\Omega_{cdm}^{} h^2=0.111(6)$\,
\cite{Beringer:1900zz}. Thus in the case of $m_{Z'}\gg m_{Z^0}$, the
DM relic density is superabundant. Therefore this scenario is not
consistent with both the LEP data and the observed DM relic density,
so that should be abandoned.

Below we turn to another possibility that $m_{Z'}\ll m_{Z^0}$.

\subsubsection{The case $m_{Z'}\ll m_{Z^0}$}

\begin{figure}[t]
\includegraphics[width=4.5in]{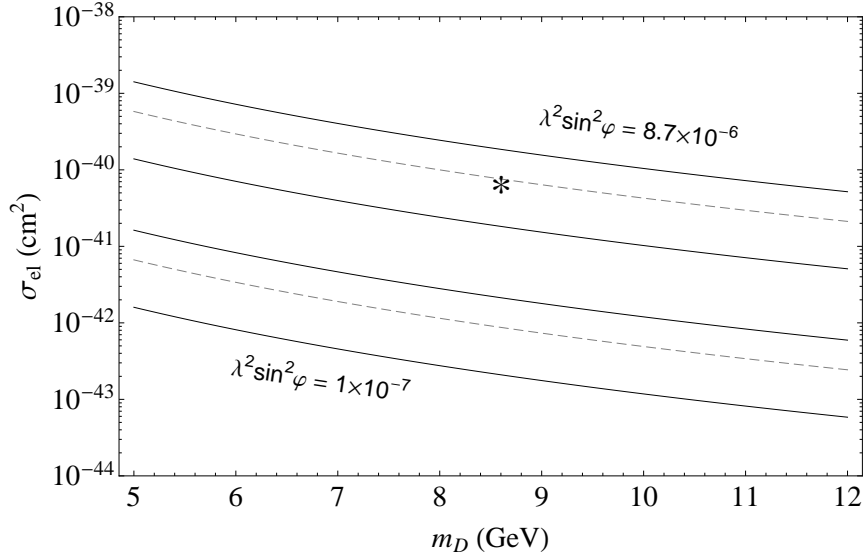} \vspace*{-1ex}
\caption{Darkon-neutron SI elastic cross-section $\sigma_{\rm
el}^{}$ as a function of the darkon's mass. $m_D$ varies in a range
$5{\rm\,GeV}\le m_D^{}\le12{\rm\,GeV}$. $2m_D^{}/m_{Z'}=\xi$, for
$\xi$ equal to 0.7, 1, 1.25.
$\lambda^2\,\sin^2\varphi=8.7\times10^{-6},1\times10^{-7}$. The
dashed curve is in the case $\xi=$1, the upper solid curve is
$\xi=$1.25, and the the lower solid curve is $\xi=$0.7. The $\ast$
is the reserved CDMS\,II observed event. \label{Ecs}}
\end{figure}

Now, let us consider the case of $m_{Z'}\ll m_{Z^0}$. If the pole
mass of $Z'$ is just slightly above $2m_D$, the annihilation cross
section of darkon pair can be enhanced. The darkon-nucleon elastic
scattering occurs mainly via exchanging $Z'$ in this case, and the
cross section is similar to the case for $m_{Z'}\gg m_{Z^0}$ and can
be re-written as
\begin{eqnarray}
\sigma_{\rm el} \simeq G_F\frac{m_{z^0}^2}{m_{z'}^2}\frac{\sqrt{2}
\lambda^2\,\sin^2\varphi\,m_{D}^2\,m_N^2\,(3\,{G_A^z}^2+{F_1^z}^2)}{4
\pi\,(p_{D}+p_{N})^2\, m_{z'}^2} \,\,.
\end{eqnarray}
Taking the CDMS II results for darkon-neutron elastic scattering as
our inputs, we get
$\lambda^2\,\sin^2\varphi\approx6.88\times10^{-3}\times({m_{z'}^4}/{m_{z^0}^4})$.
As $\cos\varphi\sim1$, the width of $Z'$ decaying into a darkon pair
is
\begin{eqnarray}
\Gamma_{D}'\simeq\frac{\lambda^2(m_{z'}^2-m_D^2)}{24 \pi
m_{z'}}\sqrt{1-\frac{4 m_D^2}{m_{z'}^2}}.
\end{eqnarray}
For the LEP constraint, using formula (\ref {widthZ}) and rewriting
formula (\ref {TotalZ}), we can obtain that when
$({m_{z'}^4}/{m_{z^0}^4})<0.167$, the width of $Z^0$ decaying into
neutrinos plus darkons is within the experimental tolerance
range. This can be satisfied when $Z'$ is lighter than half of the
$Z^0$ mass.

\begin{figure}[t] \vspace*{2ex}
\includegraphics[width=4.8in]{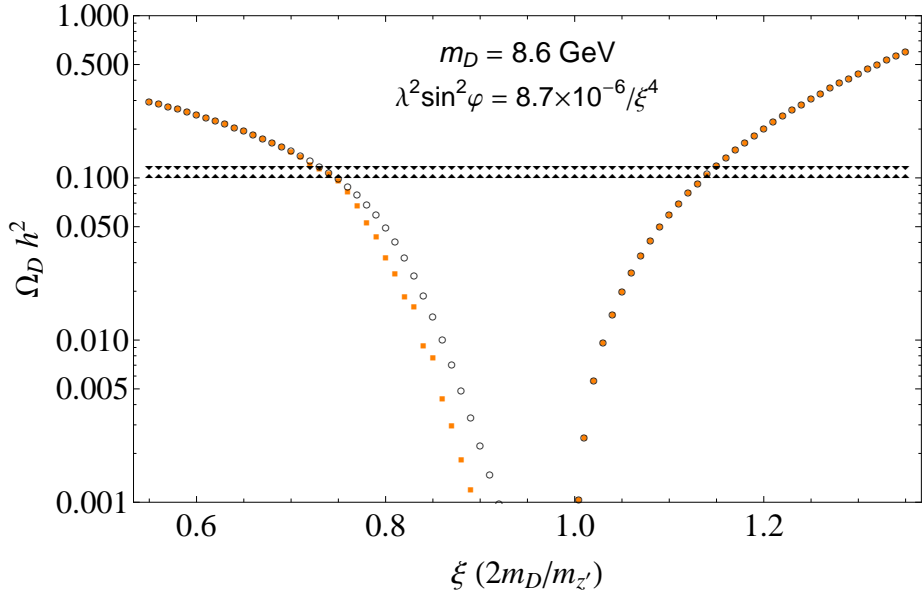} \vspace*{-1ex}
\caption{Darkon relic density $\Omega_D^{} h^2$ as a function of
$\xi$ ($2m_D^{}/m_{Z'}$) near the $Z'$ pole when $m_D=$ 8.6 GeV, for
$\xi$ in a range from 0.55 to 1.35 and
$\lambda^2\,\sin^2\varphi=8.7\times10^{-6} / \xi^4$. The solid
square curve is in the case $\lambda=0.5$, and the empty dotted
curve is $\lambda=$1.0. The triangle and triangle-down curves are
the cold dark matter relic density 0.111(6) today.}\label{Rd8.6}
\end{figure}

At the leading order, the annihilation of a darkon pair into SM
particles is determined by $Z'$ with the mixing component and the
cross section is calculated by formula (\ref {cs-general}). When
$m_{Z'}<2m_D$, the annihilation of a darkon pair into SM particles
can also pass the constraints set by the aforementioned collider
experiment and astronomical observation.

Define $2m_D^{}/m_{Z'}=\xi$. By fitting the data, in the case the
$Z'$ mass is near $2m_D$, we obtain
$\lambda^2\,\sin^2\varphi\simeq8.7\times10^{-6}$ ($\xi=1$) in the
darkon-neutron SI elastic cross-section. The dependence of the
elastic scatting cross section on $m_D$ is shown in Fig.~\ref{Ecs},
where $m_D$ varies within a range of $5{\rm\,GeV}\le
m_D^{}\le12{\rm\,GeV}$ and $\xi$ takes the values of 0.7, 1, 1.25.
$\lambda^2\,\sin^2\varphi=1\times10^{-7}$ is given as a comparison.
For $m_D\sim8.6$ GeV, fitting the results of CDMS, we have
$\lambda^2\,\sin^2\varphi\simeq8.7\times10^{-6}/{\xi}^4$.

The dependence of the darkon relic density $\Omega_D^{} h^2$ on
$\xi$ ($2m_D^{}/m_{Z'}$) is depicted in Fig.~\ref{Rd8.6} where $m_D$
is set to be 8.6 GeV and $\xi$ varies from 0.55 to 1.35.
$\lambda^2\,\sin^2\varphi\simeq8.7\times10^{-6}/{\xi}^4$. The solid
square, empty dotted curves are for $\lambda=$0.5, 1.0 respectively.
When $\xi>1$, the curve for $\lambda=$0.5 is close to the curve for
$\lambda=$1.0. It can be seen that, there is a parameter space
allowed by the present data.

\begin{figure}[t] \vspace*{2ex}
\includegraphics[width=4.8in]{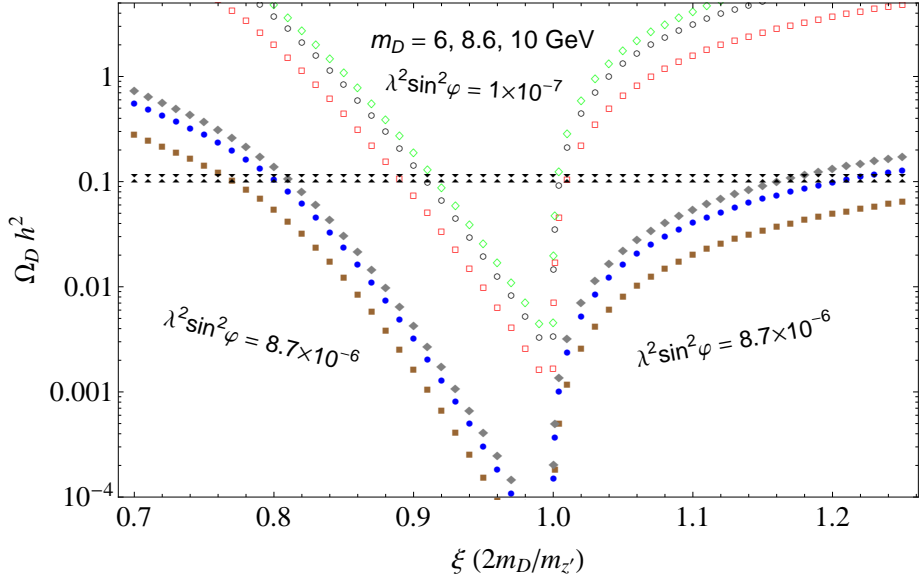} \vspace*{-1ex}
\caption{Darkon relic density $\Omega_D^{} h^2$ as a function of
$\xi$ ($2m_D^{}/m_{Z'}$) near the $Z'$ pole when $m_D=$ 6, 8.6, 10
GeV, for $\xi$'s values varying from 0.7 to 1.25. $\lambda=1$ is
taken here. The solid curves are for the case
$\lambda^2\,\sin^2\varphi=8.7\times10^{-6}$, and empty curves are
for the case $\lambda^2\,\sin^2\varphi=1\times10^{-7}$. The square
curves, dotted curves and diamond curves (solid, empty) are
corresponds to the case $m_D$ equal to 6, 8.6 and 10 GeV
respectively. The triangle and triangle-down curves are the cold
dark matter relic density 0.111(6).}\label{Rdt}
\end{figure}

As a comparison, the dependence of the darkon relic density
$\Omega_D^{} h^2$ on $m_D$ and $\lambda^2\,\sin^2\varphi$ is
shown in Fig.~\ref{Rdt} where $m_D$ is set as 6, 8.6, 10 GeV and
$\lambda^2\,\sin^2\varphi=8.7\times10^{-6}$,$1\times10^{-7}$. We take
$\lambda=1$ here and let $\xi$ vary from 0.7 to 1.25. The solid curves,
empty curves are for the case
$\lambda^2\,\sin^2\varphi=8.7\times10^{-6}$, $1\times10^{-7}$
respectively. The square curves, dotted curves and diamond curves
are corresponding to the case $m_D$ equal to 6, 8.6 and 10 GeV
respectively.

\subsection{Scalar and vector darkons}

\begin{figure}[t]
\includegraphics[bb=194 653 375 787,width=1.6in]{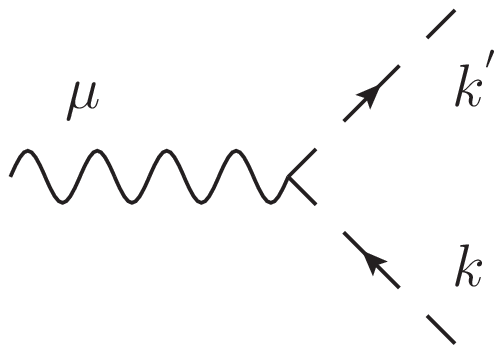} \vspace*{-1ex}\,
\, \,
\includegraphics[bb=181 607 382 784,width=1.5in]{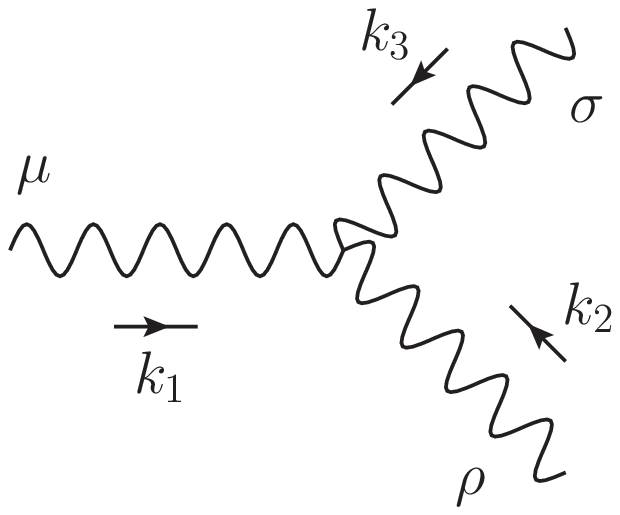} \vspace*{-1ex}
\caption{Vertexes of scalar (left) and vector (right) darkons.
}\label{scalarv}
\end{figure}

Now let us consider the scalar-darkon case. The effective vertex is
a vector coupling $-i \lambda (k+k')^{\mu}$, as shown in
Fig.~\ref{scalarv} (left). As aforementioned, the scattering of
darkon-nucleon scattering induced by this interaction is a
unsuppressed SI process \cite{Freytsis:2010ne}. In the limit
$\frac{P^\mu}{m}\rightarrow(1,\epsilon)$, the darkon-nucleon elastic
scattering cross section by exchanging $Z^0$ is written as
\begin{eqnarray}
\sigma_{\rm el} \simeq \frac{\sqrt{2}
G_F\lambda^2\,\sin^2\varphi\,m_{D}^2\,m_N^2\,{F_1^z}^2}{\pi\,(p_{D}+p_{N})^2\,
m_{Z^0}^2} \,.\label{scalarf}
\end{eqnarray}
$F_1^z=$ $0.5-2\sin^2\theta_w$ ($-0.5$) for proton (neutron). Thus,
the darkon-neutron scattering is dominant. Similar results can be
obtained as for the fermionic darkon case.

Instead, for the case that $Z'$ exchanging is dominant, one can
modify the formula (\ref{scalarf}) by simply multiplying a factor
${m_{z^0}^4}/{m_{z'}^4}$.

For the vector darkon, the vertex is $-i \lambda
[g^{\mu\rho}(k_2-k_1)^{\sigma}+g^{\rho\sigma}(k_3-k_2)^{\mu}+g^{\sigma\mu}(k_1-k_3)^{\rho}]$,
corresponding to the effective interaction
$B_{\mu}^{\dagger}\partial^{\nu}B^{\mu}\bar q\gamma_{\nu}q$, as
shown in Fig.~\ref{scalarv} (right) which contributes an unsuppressed
SI cross section. In the limit
$\frac{P^\mu}{m}\rightarrow(1,\epsilon)$, the darkon-nucleon elastic
scattering cross section with $Z^0$ exchange-dominance can be
written as
\begin{eqnarray}
\sigma_{\rm el} \simeq \frac{\sqrt{2}
G_F\lambda^2\,\sin^2\varphi\,m_{D}^2\,m_N^2\,{F_1^z}^2}{\pi\,(p_{D}+p_{N})^2\,
m_{Z^0}^2} \,\label{vectorf}.
\end{eqnarray}

In the case that $Z'$ exchange is dominant, the elastic cross
section can be obtained by multiplying  formula (\ref{vectorf}) by a
factor ${m_{z^0}^4}/{m_{z'}^4}$. The obtained result is the same as that
for scalar-darkon-nucleon elastic scattering.

It is noted, for fermionic, scalar and vector darkon-nucleon elastic
scattering via exchanging $Z-$boson, there exists SI darkon-neutron
scattering which are not suppressed by either $q^2$ or $v^2$. In
this case, the proton contributions are suppressed, so that the main
contributions to the SI scattering comes from the interaction
between the darkon and neutron. Therefore, the xenon target which
has more neutrons than protons is more sensitive compared with the
silicon and germanium targets. As explained above, if we accept the
claim of XENON10~\cite{Angle:2011th} and
XENON100~\cite{Aprile:2012nq} that for low energy WIMPs, null
results have been obtained, the CDMS results should be dubious. But
suggested by Hooper \cite{Hooper:2013cwa}, a re-analysis may imply
that the peculiar events observed at the XENON100 might be explained
as dark matter candidates to be reconciled with the CDMS data. With
the further progress of the XENON experiments, more information will
be obtained for the low mass WIMPs with masses of order 10 GeV.

\section{Conclusions and discussion}

Taking the recent new results of the CDMS II experiments searching
for WIMPs with masses of order 10 GeV as inputs and considering some
constraints from LHC, LEP and astronomical observation etc.
altogether, we discuss a simple WIMP candidate: the darkon which can
be scalar, fermion or vector. We have found that in the simplest
scenario of the standard model plus a SM singlet DM (the darkon),
one cannot simultaneously satisfy the CDMS II's observation and the
LHC data, and this result is consistent with the former result
implied in~\cite{He:2007tt}. Thus, one must extend the SM to include
new physics beyond the standard model. Here we consider the extended
gauge group $SU_L(2)\otimes U_Y(1)\otimes U(1)'$ which later breaks
into $U_{em}(1)$ to result in two heavy neutral gauge bosons $Z^0$
and $Z'$.

The darkon+SM+$U(1)'$ scenario must undergo stringent tests from the
cosmology observation and the LHC data. Namely, all the CDMS II
results, dark matter density in our universe and the data of $Z^0$
decaying into invisible products which were obtained by LEP
experiments must not conflict.

Our numerical results indicate that in this scenario, only if $Z'$
is lighter than $Z^0$, all the constraints can be satisfied. Under
this assumption, the model darkon+SM+$U(1)'$ withstand all the
constraints set by the presently available data. Moreover, it is noted
that as long as  $m_{Z'}\sim 2m_D^{}$, the model can accommodate
even smaller scattering cross section and lighter darkons.

Indeed we should further test the validity of this mechanism. If in
the future, we can precisely measure the branching ratios of heavy
quarkonia, such as botomonia decaying into invisible products, or
the invisible decays of the SM Z boson and Higgs sector physics, we
would be able to determine which one e.g., of the two-Higgs-doublets
mechanism or the extra $U(1)'$ gauge group, is more reasonable. We
lay hope on the future more precise detection on the dark matter, no
matter direct or indirect.

In the world today there are many laboratories directly searching
for dark matter besides the XENON and CDMS collaborations, for
example, the China Jin-Ping underground laboratory
\cite{Kang:2013sjq} just joined the club and the China Dark-Matter
experiment (CDEX) is using 1 kg Ge detector and will develop 10 kg
and 1 ton detector for the project. We are expecting that the
world-wide cooperation can eventually reveal the Epoch mystery.

\acknowledgments \vspace*{-3ex} We thank Prof. X.G. He for helpful discussions. This work was partially supported by
National Natural Science Foundation of China  under the contract No.11075079, 11135009.

\end{document}